\documentclass{aa}  

\usepackage{graphicx}
\usepackage{txfonts}
\usepackage{lipsum}
\usepackage{subcaption}         % necessary for continued figures, example in section 3
\usepackage{lscape}             % to rotate a single page table, example in appendix.
\usepackage{placeins}           % useful with \FloatBarrier, to keep 
\usepackage{tabularx}

\begin{document}
%%%%%%%%%%%%%%%%%%%%%%%%%%%%%%%%%%%%%%%%
% if you use custom commands in your title,
% ensure to check your title when submitting!
%%%%%%%%%%%%%%%%%%%%%%%%%%%%%%%%%%%%%%%%
\title{Predicted Ejecta Dynamics and Observability of the 2026 Falcon 9 Upper Stage Lunar Impact}
%%%%%%%%%%%%%%%%%%%%%%%%%%%%%%%%%%%%%%%%
% Please separate each author with the \and command
%
% Use the \corrauth to provide the corresponding
% author address. It will be automatically inserted as 
% footnote in the PDF output.
%
% Please DO NOT include ORCIDs next to author names.
% Instead, please provide an active address for each coauthor:
% it will be automatically extracted by EDPS editorial system, 
% and co-authors will be be able to authenticate their ORCID.
%
% Only authenticated ORCIDs will be taken into account.
% ORCIDs included here will be removed.
%%%%%%%%%%%%%%%%%%%%%%%%%%%%%%%%%%%%%%%%

\author{William Jo\inst{1}\corrauth{williamjo@utexas.edu}   
\and David B. Goldstein\inst{1}\email{fab@nestor-edp.org}
\and Philip L. Varghese\inst{1}\email{cat@nestor-edp.org}
\and Laurence M. Trafton\inst{2}\email{clem@nestor-edp.org}
\and Jordan Steckloff\inst{1,3}\email{lolo@nestor-edp.org}
\and Arnaud Mahieux\inst{1,4,5}\email{narg@nestor-edp.org}}

\institute{Department of Aerospace Engineering and Engineering Mechanics, The University of Texas at Austin, Austin, TX
\and Department of Astronomy, The University of Texas at Austin, Austin, TX
\and Planetary Science Institute, Tucson, AZ
\and Belgian Royal Institute for Space Aeronomy, Brussels, Belgium
\and SSC Space for the European Space Agency (ESAC), Madrid, Spain}

\date{Received August 4th, 2026}

\abstract
{On Aug.~5 at 06:34~UTC, a Falcon~9 upper stage ($\sim$3900\,kg) will strike the lunar surface at 2.43\,km/s, yielding a potentially visible debris plume. We present a study of the expected impact dynamics and resulting possibly observable debris field. The debris plume should reach roughly 15 to 20\,km in altitude for the ejecta curtain and 75 to 100\,km for the central ejecta spike, extend 183\,km laterally from the impact point near the sunlit limb, and yield a peak dust column density above 10\,km altitude of $6.08 \times 10^{7}~\mathrm{m^{-2}}$. Simulated $I/F$ exceeds dark-sky background brightness by several orders of magnitude for the first several minutes after impact, reaching $I/F \approx 1.27 \times 10^{-3}$ at the earliest resolved time ($t = 5$\,s). Above 10\,km, peak $I/F$ reaches $1.33 \times 10^{-5}$, still several orders of magnitude brighter than the dark sky.}


\maketitle
\nolinenumbers

%%%%%%%%%%%%%%%%%%%%%%%%%%%%%%%%%%%%%%%%%%%%%%%%%%%%%%%%%%%%%%
\section{Introduction}
{Spacecraft and spacecraft-fragment impacts on the Moon have been observed since the Ranger and Luna missions of the 1960s. Unlike natural meteoroid impacts, these events occur under well-constrained conditions, making them valuable for studying lunar properties and impact physics. During the Apollo era, controlled impacts provided seismic sources that probed the lunar interior \citep{nakamura1982apollo}. More recently, the LCROSS mission successfully identified volatile species, including water and water ice, in the permanently shadowed regions of the lunar south pole \citep{colaprete2010detection}. In contrast, an earlier attempt to detect water or OH released by the Lunar Prospector impact was unsuccessful \citep{goldstein1999impacting}.
    
 Whereas meteorites generally strike the lunar surface with a median speed of $\sim$15.9 km/s \citep{yue2013projectile}, spacecraft strike the surface at $\sim$2 to 3 km/s \citep{colaprete2010detection, plescia2016ranger}. Thus, while meteorite impact physics are described in terms of structural deformation, melting, vaporization, chemistry, and even ionization of the impactor and regolith, spacecraft impacts are gross fragmentation and cratering phenomena, with little relevance to melting, vaporization, chemistry, ionization and bright flashing \citep{ahrens1972shock}. A second important difference between Moon impacts of meteorites and spacecraft is that for modeling spacecraft impacts, the internal structural details are relevant to the debris plume produced \citep{hermalyn2012scouring, rajsic2021numerical, owen2022spacecraft}. Assuming little residual fuel, the impacting mass of the Falcon 9 upper stage spacecraft will reside largely in the spacecraft shell, tanks and a relatively compact dense engine. Natural impactors, in contrast, are commonly modeled as a more uniform density and regular shape, possibly with an assumed porosity distribution or dense nodules or inclusions: they are never considered as thin hollow shells. Here we present iSALE-2D simulations \citep{amsden1980sale, collins2004modeling, wunnemann2006strain}, of the anticipated vertical, engine-first impact, coupled to a Monte Carlo ballistic reconstruction of the three-dimensional ejecta plume and its predicted brightness in the form of $I/F$, the ratio of the radiance of the plume particles to the solar flux at the location of Earth.}

\begin{figure*}
    \centering
    \includegraphics[width=0.9\textwidth]{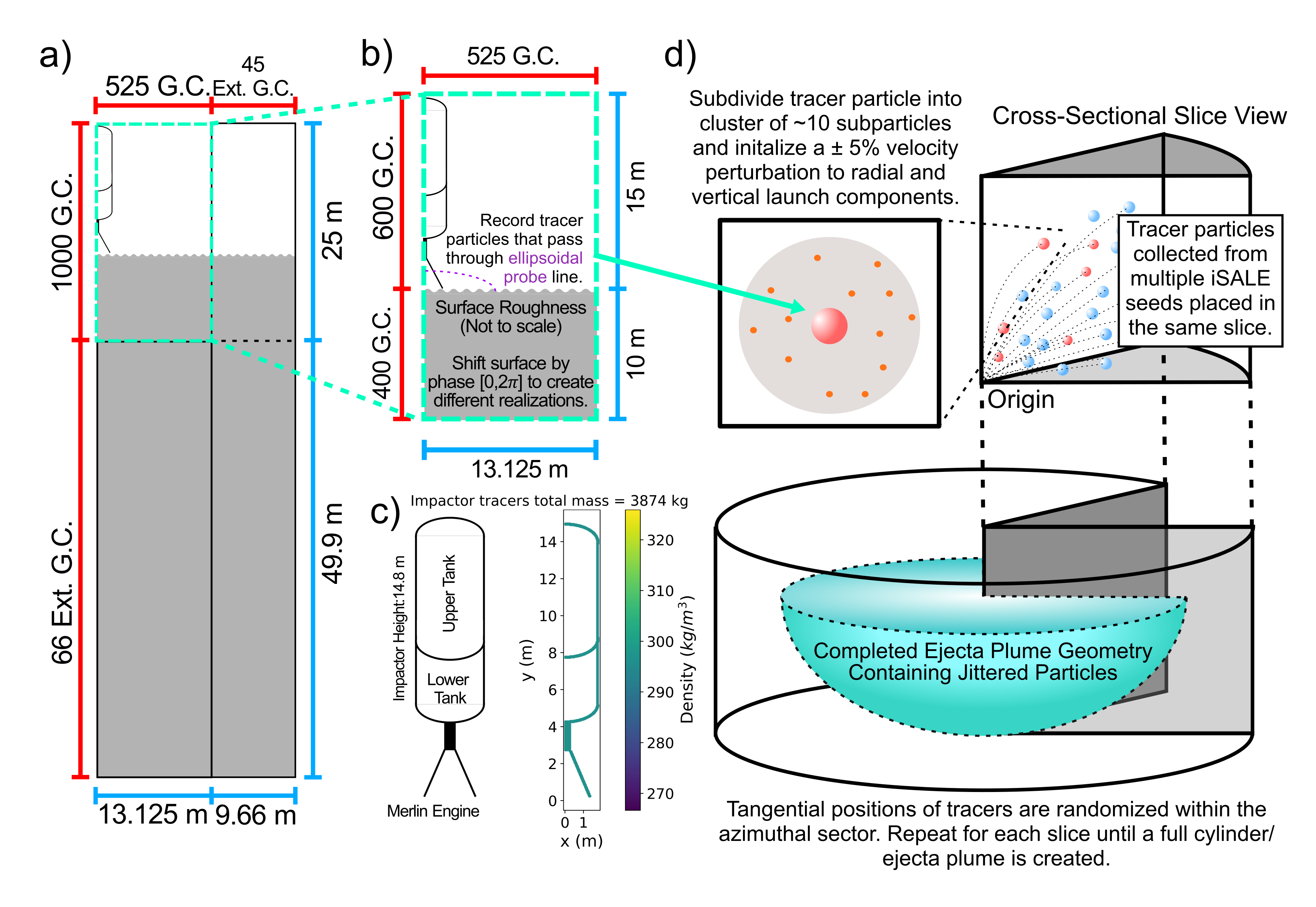}
    \caption{iSALE-2D computational domain and Falcon~9 upper stage impactor geometry. (a) Full axisymmetric domain. (b) Structural wall resolved four cells across a 10~cm thickened shell. (c) Impactor construction as six solid shapes and five voided regions (upper and lower propellant tanks and Merlin engine). (d) Sub-particle jittering scheme applied to each tracer crossing the hemi-ellipsoidal probe.}
    \label{fig:domain_geometry}
\end{figure*}

\section{Materials and Methods}

{The iSALE computational domain and Falcon 9 impactor geometry (upper and lower propellant tanks and Merlin engine, modeled as six solid shapes and five voided regions) are described in Figure~\ref{fig:domain_geometry}a to c. The model of the Falcon 9 upper stage is derived from \citet{wittal2022ambiguity} with the impact speed set to 2.43 km/s \citep{gray2026upper}. We assume the impactor has exhausted its propellant before impact, so that any residual propellant does not affect plume formation. We also assume that the rocket body strikes the surface vertically enabling an axisymmetric simulation. More details about the construction of the impactor on iSALE are available in Appendix \ref{app:geometry}. The high-resolution region of the computational mesh uses square cells of 2.5 cm, and the Falcon 9 upper stage's outer shell is resolved as four cells across, corresponding to a 10 cm thick structural wall, as described in Figure 1b. Because the true wall thickness of the Falcon 9 impactor ($\sim$4.7 mm) is far below the grid resolution, the shell cannot be modeled directly. Instead, we raise the wall's porosity to approximately 95$\%$, so that the mass per unit area of the thickened (10 cm) wall matches the actual Falcon 9 upper stage dry vehicle mass of $\sim$3,900 kg \citep{wevolver2026falcon9}. This approach follows prior iSALE work in which impactor porosity is raised to mimic the hollowness of the Apollo S-IVB upper stage in lunar impact simulations \citep{rajsic2021numerical}.

The stainless-steel material used for the impactor was represented with a modified Tillotson equation of state (EOS) for iron \citep{tillotson1962metallic, maurel2020simulations}. The lunar regolith was defined using a modified basalt EOS \citep{benz1999catastrophic, wojcicka2020seismic, rajsic2021numerical}. The lunar target is represented as a five-layer porous regolith column, with layer porosities assigned from the depth-dependent highland porosity profile of \citet{carrier1991physical}. Ejected tracer mass is converted into particle counts using the Apollo 16 grain-size distribution, also from \citet{carrier1991physical}. The distribution spans grain diameters from 1 µm to 1.6 mm across 10 bins, with the largest mass fractions concentrated in the 40 to 84 µm range. The number of particles in each bin follows from the bin mass and the mass of an individual grain at the bin's mean diameter. The specific input parameters, EOS values, and bin distributions for these models are provided in Appendix \ref{app:eos} to \ref{app:psd}.
    
In iSALE's 2D-axisymmetric formulation, ejecta tracers are emitted as discrete mass rings rather than point particles, so a single simulation does not resolve individual particle distributions and yields a spatially noisy ejecta field, whereas a physical plume is a more continuous density field. To convert the ring-averaged tracer output into a 3D particle distribution, we use a Monte Carlo particle method that takes each tracer's mass and velocity as input for 3D post-processing. Under the assumption of Moon's low-gravity, near-vacuum environment, ejecta decouples from the impact flow near the crater and subsequently follows ballistic trajectories \citep{melosh1989impact}.

A further complication is that a single simulation provides limited statistics for the fastest-moving material. These are the particles of greatest interest here, since these are the ones observable farthest off the lunar surface. Early ejecta are the fastest and reach the highest altitudes but are poorly sampled because most ejecta mass is released at lower velocities. To improve statistics for this high-altitude population, we perform an ensemble of impact simulations with low-amplitude sinusoidal regolith surface roughness based on \citet{guo2021millimeter}, varying its phase between simulations. Each phase shift leaves the bulk impact outcome unchanged while introducing stochasticity into individual ejecta trajectories. In total, 16 unique simulations are generated by phase-shifting a sinusoidal surface roughness with wavelength 1.37~m and peak-to-peak amplitude 10~cm from 0 to $2\pi$ radians. Tracer particles crossing a hemi-ellipsoidal surface just above the ground are recorded up to 25~ms after the impactor's contact with the surface.

While phase-shifting the surface roughness introduces stochasticity between simulations, the particles within a single tracer's ring remain otherwise identical, and still land on a shared azimuthal locus rather than form a dispersed cluster. To break this residual uniformity, we further jitter each tracer at the sub-particle level, as described in Figure~\ref{fig:domain_geometry}d. A tracer crossing the hemi-ellipsoidal probe is therefore replaced by roughly a dozen sub-particles whose masses sum to that of the original tracer and which inherit its launch velocity. Each sub-particle's radial and vertical velocity components are then perturbed by up to $\pm5\%$ to introduce additional stochasticity, such that the cluster disperses along slightly diverging ballistic paths. Even after subdivision, the tracer count in a single simulation remains too sparse to represent a continuous plume, and each realization describes only a single 2D slice. Reconstruction of the third dimension exploits the fact that each phase-shifted simulation is an equally valid sample of the same impact. The radial and vertical states of the subparticles are retained, while their azimuthal positions are assigned randomly within a sector, and the process is repeated sector by sector around the full circumference until the plume volume is populated. A total of 32 slices are used to simulate the impact, where each slice is $\pi/16$ radians in angular width.

The particle column density of the ejecta plume is then computed by binning subparticles into a fixed 3D grid spanning $\pm800$~km in the horizontal ($x, z$) directions and $\pm200$~km in the vertical ($y$) direction, with 5~km resolution in each dimension. At each timestep, we accumulate grain counts per voxel, convert to number density by dividing by voxel volume, then integrate along the vertical axis to yield a two-dimensional column density map in $\mathrm{grains/m^2}$. To connect this to observable quantities, we post-process the two-dimensional column density maps into $I/F$, defined as the ratio of measured radiance of the ejecta plume to the solar flux. 

For scattering by a perfectly spherical particle, we first obtain the scattering matrix relating the incident and scattered electric fields (Equation~\ref{eq:smatrix}), where $S_1$ and $S_2$ are the Mie scattering amplitude functions. Mie scattering matrices and cross sections are computed with MatScat \citep{bohren2008absorption, schafer2011implementierung, schaefer2012calculation}:
\begin{equation}
\label{eq:smatrix}
   \begin{bmatrix} E_{\parallel s} \\ E_{\perp s} \end{bmatrix}
   = \frac{\exp[ik(r-z)]}{-ikr}
   \begin{bmatrix} S_2 & 0 \\ 0 & S_1 \end{bmatrix}
   \begin{bmatrix} E_{\parallel i} \\ E_{\perp i} \end{bmatrix} \,.
\end{equation}

From the amplitude functions and the wavenumber $k$, the differential scattering cross section follows, and the phase function $P(\theta)$ in turn from the scattering cross section $C_{\mathrm{sca}}$ \citep{draine2013user, draine2003scattering}, with resulting phase function shown in Figure~\ref{fig:phase_function}. To compute the phase function, we assume the regolith complex refractive index is taken as the lunar simulant JSC-1A (m = 1.65 + 0.003i) \citep{goguen2010new}.  For a sample observer such as the Hubble Space Telescope (HST), the scattering angle at the date and time of the impact is $\theta_{sc} = 99.9^{\circ}$ \citep{giorgini1996jpl}. The JPL Horizons calculation used to determine this viewing geometry is summarized in Appendix~\ref{app:geometry_sto}.

\begin{equation}
\label{eq:dcsca}
   \frac{\mathrm{d}C_{\mathrm{sca}}}{\mathrm{d}\Omega}
      = \frac{1}{2k^2}\left(|S_1|^2 + |S_2|^2\right) \,.
\end{equation}

\begin{equation}
\label{eq:phase}
   P(\theta)
      = \frac{1}{C_{\mathrm{sca}}}\,
        \frac{\mathrm{d}C_{\mathrm{sca}}}{\mathrm{d}\Omega} \,.
\end{equation}

\begin{figure}[!t]
    \centering
    \includegraphics[width=\columnwidth]{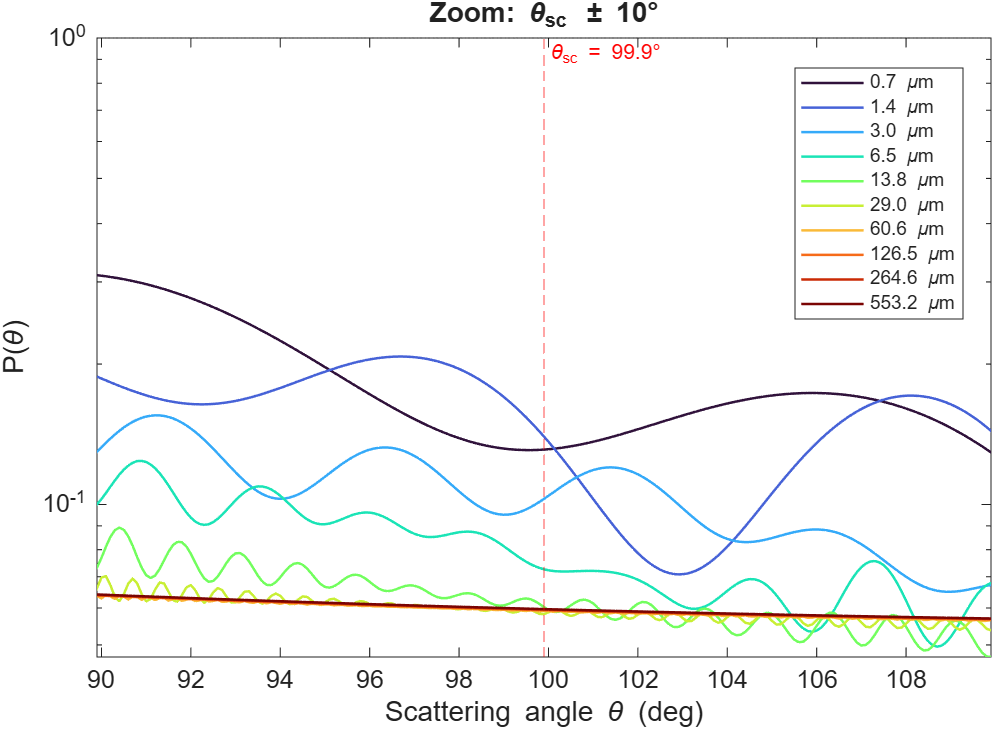}
    \caption{Mie scattering phase function $P(\theta)$ for regolith grains with index of refraction $m = 1.65 + 0.003i$ across the ten Apollo~16 PSD size bins, at $\lambda = 550$~nm, post-processed from the MatScat program \citep{bohren2008absorption, schafer2011implementierung, schaefer2012calculation}, with the red line marking the scattering angle $\theta_{sc} = 99.9^{\circ}$.}
    \label{fig:phase_function}
\end{figure}

We then evaluate an observable $I/F$, the ratio of the measured scattered intensity $I$ to the flux of a perfectly Lambertian white surface illuminated perpendicularly by the Sun ($\pi F$), and sum the contributions of the 10 particle size bins indexed by $j$ (Equations~\ref{eq:if} and~\ref{eq:iftotal}), where $C_{\mathrm{sca}}$ and $C_{\mathrm{ext}}$ are the single-particle scattering and extinction cross sections and $N_{\mathrm{los}}$ is the line-of-sight column density \citep{seignovert2017aerosols}. We repeat this for every column density grid cell per timestep.
\begin{eqnarray}
\label{eq:if}
   I/F & = & \frac{C_{\mathrm{sca}}}{C_{\mathrm{ext}}} \cdot
             \frac{P(\theta)}{4} \cdot
             \left[1 - \exp\left(-N_{\mathrm{los}}\,C_{\mathrm{ext}}\right)\right]
\end{eqnarray}
\begin{eqnarray}
\label{eq:iftotal}
   (I/F)_{\mathrm{total}} & = & \sum_{j=1}^{10} (I/F)_j \,.
\end{eqnarray}

From the Pogson equation, relating a magnitude difference $(\mu_2 - \mu_1)$ to a flux ratio $(F_2/F_1 = I/F)$, the apparent surface brightness of the ejecta $\mu_2$ follows; applying it a second time with the Sun as reference fixes $\mu_1$ through the geometric flux ratio:
\begin{equation}
\label{eq:pogson1}
   \mu_2 = \mu_1 - 2.5\,\log_{10}(I/F) \,.
\end{equation}

Here, $\mu_1$ is the surface brightness of a perfectly diffuse (Lambertian) reference surface. To find this value, we apply the Pogson equation a second time, now setting the reference to the Sun itself, as shown in Equation~\ref{eq:pogson2}. Here, $\mu_0 = -26.74$, is the apparent magnitude of the Sun, $F_0$ is the solar flux at 1~AU, and $F_1$ is the flux per arcsec$^2$ from the Lambert reference plate. Note that $F_0$ and $F_1$ are both scaled with the solar flux $S_\odot$, which cancels in the ratio $F_1/F_0$, leaving only the form represented in Equation~\ref{eq:pogson2}. The values $\Omega = 1~\mathrm{arcsec}^2 = 2.3504 \times 10^{-11}~\mathrm{sr}$ is the solid angle corresponding to one square arcsecond, and $r_h$ is the heliocentric distance in AU.
\begin{equation}
\label{eq:pogson2}
   \frac{F_1}{F_0} = \frac{\Omega}{\pi r_h^2} \,.
\end{equation}

At 1~AU ($r_h = 1$), the value of $\mu_1$ becomes $1.08~\mathrm{mag/arcsec^2}$. Thus, the apparent brightness $\mu$ can be written as a function of $I/F$, as described in Equation~\ref{eq:pogson3}:
\begin{equation}
\label{eq:pogson3}
   \mu(I/F) = 1.08 - 2.5\,\log_{10}(I/F) \,.
\end{equation}
}

\section{Data and Results}
\begin{figure*}[!b]
    \centering
    \includegraphics[width=0.85\textwidth]{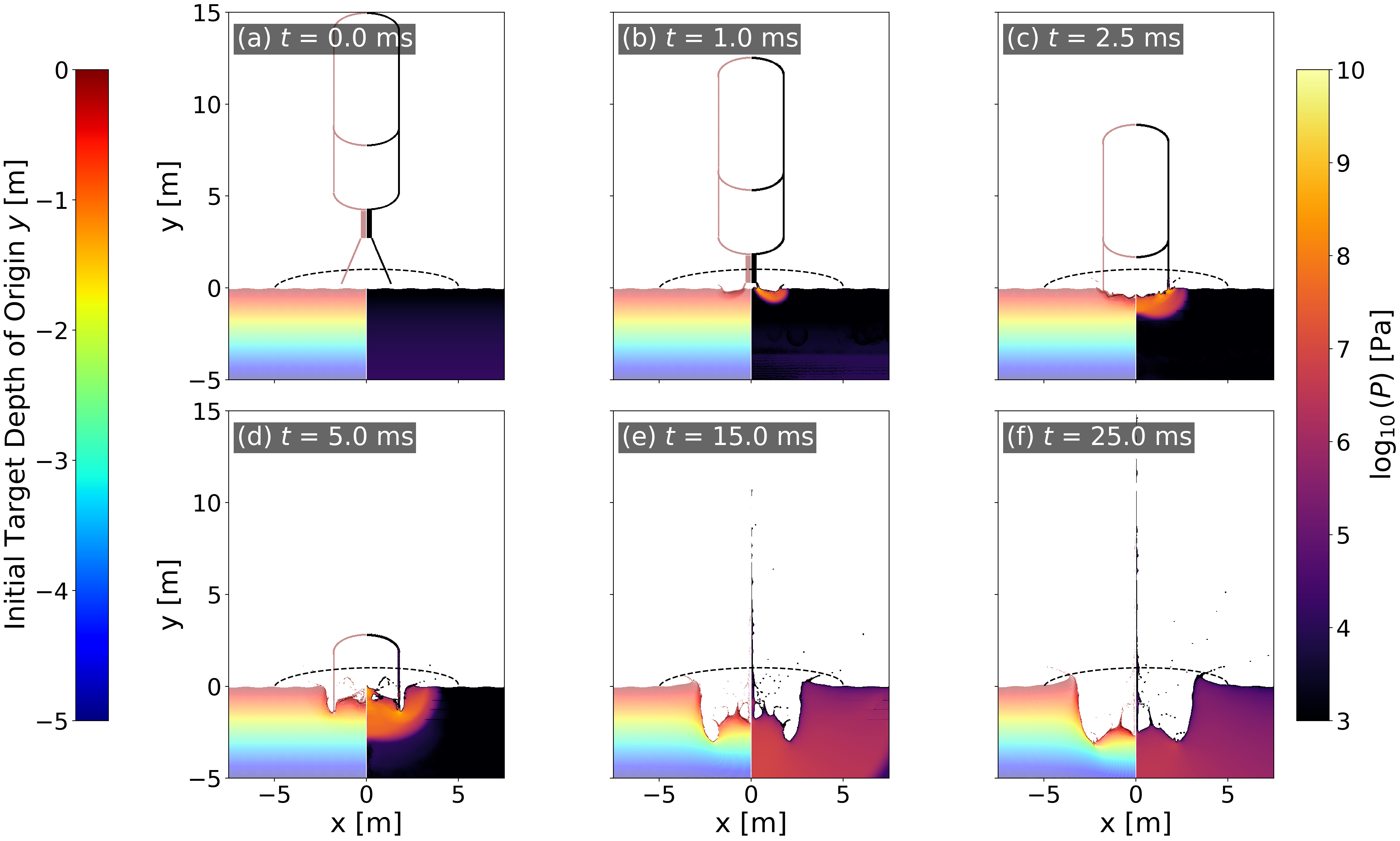}
    \caption{Impact evolution for a single nominal upright Falcon~9 configuration from 0 to 25~ms. Each panel is mirrored across the symmetry axis, with tracer particles colored by initial provenance depth (left half) and the pressure field (right half). (a)--(c) Initial penetration along the centerline as the engine disintegrates, opening an approximately parabolic transient cavity. (d) Excavation driven by the side walls once the centerline contribution stalls. (e) Formation of the near-vertical central spike as the trailing end of the impactor reaches the surface. The dashed curve marks the hemi-ellipsoidal tracer probe.}
    \label{fig:isale_impact}
\end{figure*}

Figure~\ref{fig:isale_impact} shows the impact evolution for a single nominal upright Falcon~9 configuration from 0 to 25~ms, with tracer particles colored by initial provenance depth (left half of each panel) and the pressure field (right half), mirrored across the symmetry axis. Upon first contact with the lunar target (Figures~\ref{fig:isale_impact}a to c), the impactor behaves in a manner qualitatively distinct from a solid projectile of equivalent mass. The simulated impact shows that the initial penetration is driven along the centerline as the engine is disintegrated, opening a transient cavity that is approximately parabolic in cross-section. As the impact progresses (Figure~\ref{fig:isale_impact}d), however, the collapsing structure no longer loads the target along the axis. We observe that excavation is driven almost entirely by the side walls, which continue to couple momentum into the target while the centerline contribution stalls. This behavior suggests that a hollow impactor excavates along the axis in a fundamentally different manner than an equivalent solid projectile. Notably, the central spike does not form until the trailing end of the impactor reaches the surface, at which point a narrow, near-vertical spray is released along the axis (Figure~\ref{fig:isale_impact}e). The visual needle-like appearance of this central spike is an artifact of the truly axisymmetric nature of the computation.

\begin{figure*}[!t]
    \centering
    \includegraphics[width=0.75\textwidth]{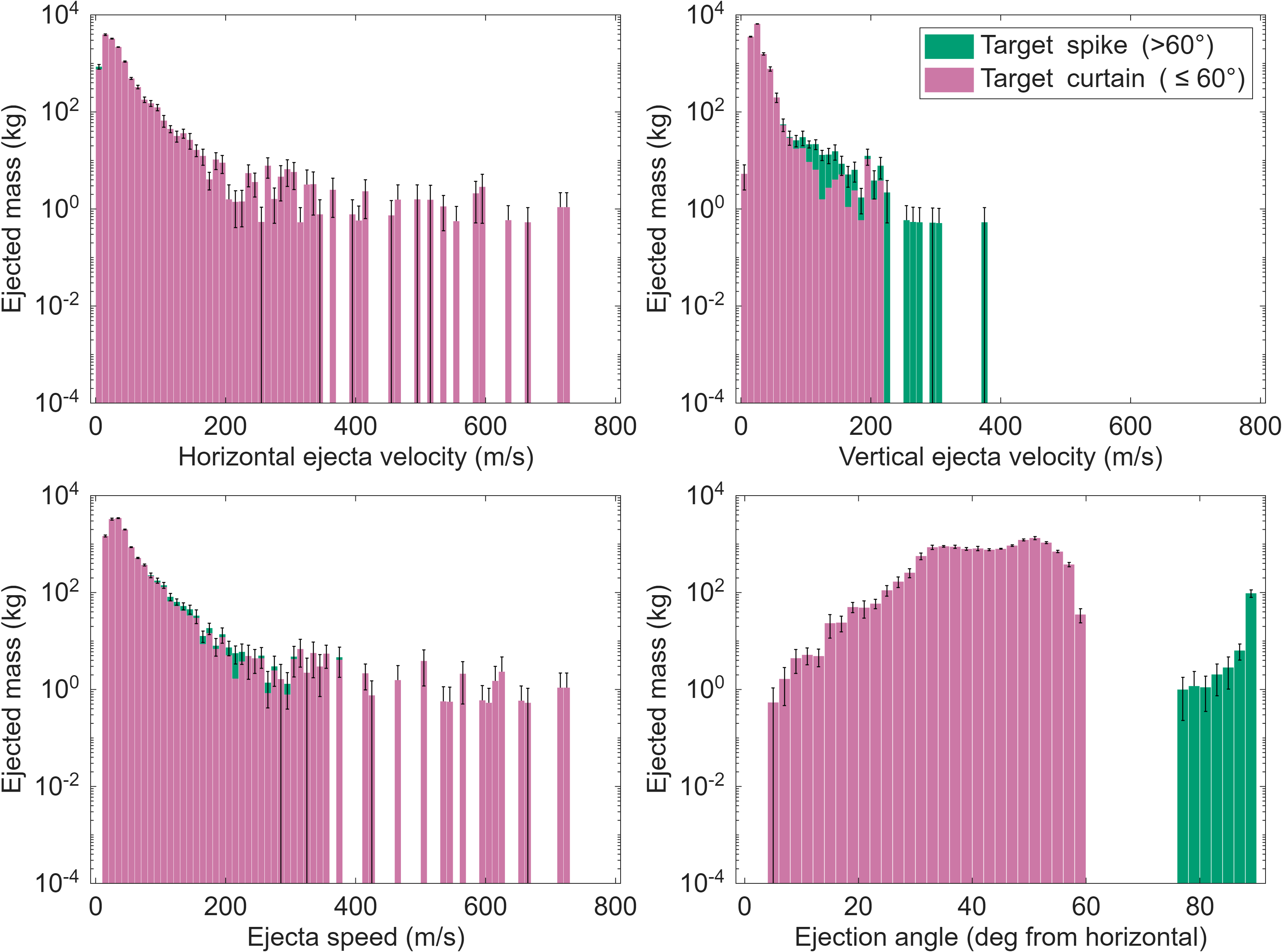}
    \caption{Mass distribution of target ejecta (spike vs.\ curtain, split at a $60^{\circ}$ ejection angle threshold) as a function of (a) horizontal ejecta velocity, (b) vertical ejecta velocity, (c) total ejecta speed, and (d) ejection angle. Bin widths are set to 10~m/s for (a) to (c) and $2^{\circ}$ for (d).}
    \label{fig:ejecta_dist}
\end{figure*}

Figure~\ref{fig:ejecta_dist} shows the mass-weighted velocity and angular distributions of the target ejecta collected by the semi-ellipsoidal tracer probe over 0 to 25~ms decomposed by initial ejection angle into curtain (0 to $60^{\circ}$) and spike (60 to $90^{\circ}$) components. Bin widths are set to 10~m/s for Figures~\ref{fig:ejecta_dist}a to c and $2^{\circ}$ for Figure~\ref{fig:ejecta_dist}d. Within this short launch timeframe, a total of $12{,}920 \pm 76$~kg is lofted beyond the dashed curve, of which the curtain carries $12{,}810 \pm 80$~kg and the spike $111 \pm 14$~kg. We find that 99.3\% of curtain mass is ejected at speeds below 200~m/s. The angular distribution (Figure~\ref{fig:ejecta_dist}d) shows curtain mass concentrated between roughly $30^{\circ}$ and $60^{\circ}$, with a sharp drop above $55^{\circ}$. The spike component appears predominantly at vertical velocities above $\sim$70~m/s (Figure~\ref{fig:ejecta_dist}b) and at ejection angles concentrated near $90^{\circ}$ (Figure~\ref{fig:ejecta_dist}d). We compute the ejecta mass remaining aloft above a series of reference altitudes, following the planetary ballistic relations of \citet{gault1963spray}. Table~\ref{tab:altitude} summarizes the minimum vertical velocity ($v_{y,\mathrm{min}}$) required for a particle to clear a given altitude under flat-plane ballistics, the curtain and spike mass exceeding that velocity, and the corresponding maximum time aloft. As expected, the curtain component depletes rapidly with altitude, falling to zero above roughly 20~km, while the central spike, despite carrying under 1\% of the total ejected mass, is expected to persist to significantly greater heights. The curtain is expected to remain aloft for less than 233~s, and the spike for less than 504~s.
\begin{table}[!t]
    \centering
    \caption{Mass clearing altitude thresholds calculated from probe-collected velocity tracers under relations used by \citet{gault1963spray}.}
    \label{tab:altitude}
    \resizebox{\columnwidth}{!}{%
    \begin{tabular}{ccccc}
    \hline\hline
    Altitude (km) & $v_{y,\mathrm{min}}$ (m/s) & Curtain mass (kg) & Spike mass (kg) & Time aloft (s) \\
    \hline
    1  & 43  & $208.7 \pm 57.3$ & $111.1 \pm 14.2$ & $< 53.0$ \\
    5  & 96  & $32.2 \pm 12.7$  & $50.2 \pm 11.4$  & $< 118$  \\
    10 & 135 & $16.7 \pm 7.2$   & $14.5 \pm 5.7$   & $< 167$  \\
    15 & 164 & $3.9 \pm 2.7$    & $6.6 \pm 3.1$    & $< 203$  \\
    20 & 189 & $0.0 \pm 0.0$    & $3.2 \pm 2.3$    & $< 233$  \\
    25 & 211 & $0.0 \pm 0.0$    & $1.6 \pm 1.1$    & $< 260$  \\
    50 & 294 & $0.0 \pm 0.0$    & $0.0 \pm 0.0$    & $< 363$  \\
    \hline
    \end{tabular}%
    }
\end{table}
    
\begin{figure*}[!t]
    \centering
    \includegraphics[width=0.75\textwidth]{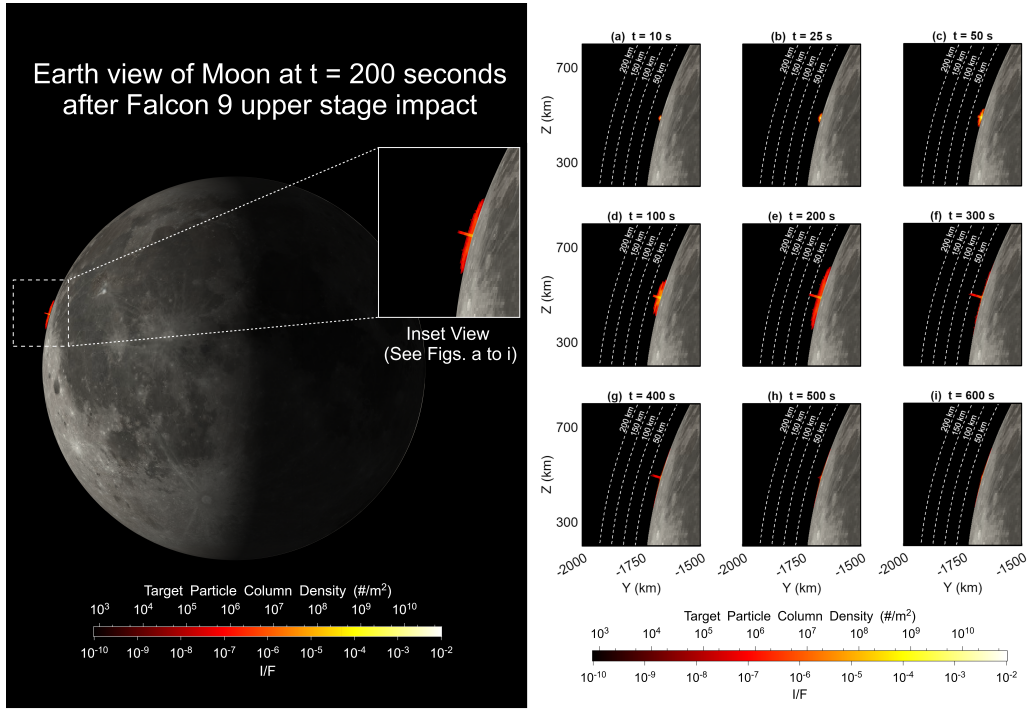}
    \caption{Global view of the Moon showing the target ejecta plume, with insets at right showing the 3D plume evolution in close-up limb view at nine representative times ($t = 10, 25, 50, 100, 200, 300, 400, 500$, and 600~s after impact), reconstructed by propagating the tracer ejection with the particle jittering method. Within the first few seconds the side-view grain column density reaches $\mathcal{O}(10^{9}~\mathrm{m^{-2}})$ near the impact point; by 100~s the plume forms a pancake-like structure with column densities of $\sim$$10^{6}~\mathrm{m^{-2}}$ and a central spike extending above it along the axis.}
    \label{fig:coldens_if}
\end{figure*}

Figure~\ref{fig:coldens_if} shows the global view of the Moon, with the insets at right showing the target ejecta plume 3D evolution in close-up limb view at nine representative times ($t = 10, 25, 50, 100, 200, 300, 400, 500$, and 600~s after impact), reconstructed by propagating the tracer ejection using the particle jittering method. Note that within the first few seconds after impact, the side view grain column density reaches $\mathcal{O}(10^{9}~\mathrm{m^{-2}})$, located near the point of impact. The plume at 100~s after impact can be described as a pancake-like structure with column densities of $\sim$$10^{6}~\mathrm{m^{-2}}$, with the spike extending above it along the axis. Above 10~km altitude, the peak column density reaches $6.08 \times 10^{7}~\mathrm{m^{-2}}$ ($6{,}080~\mathrm{cm^{-2}}$). The curtain reaches its maximum extent near 200~s, and has largely returned to the surface by 300~s. By 600~s, the spike portion of the ejecta plume has decayed out of view, and it is expected that the Falcon~9 impact event will conclude beyond this time.

We find that the peak plume brightness exceeds $I/F > 10^{-4}$ for at least $\sim$75~s after impact. At $t = 5$~s, the plume brightness reaches $I/F = 1.27 \times 10^{-3}$, as shown in Figure~\ref{fig:coldens_if}. When converted to surface brightness, the $I/F$ values for $t < 75$~s correspond to $8.32 < \mu < 11.02~\mathrm{mag/arcsec^2}$ (using $\mu = -2.5\,\log_{10}(I/F) + 1.08$). Above 10~km altitude, peak $I/F$ reaches a value of $1.33 \times 10^{-5}$ at 90~s after impact, corresponding to $\mu \approx 13.27~\mathrm{mag/arcsec^2}$. However, bright spots are very localized within 2.5~km above the surface of the Moon, rapidly diluting beyond this small bright spot. Our results suggest that although the overall plume is spatially large, only a small, bright inner region near the surface may be observable. The peak $I/F$ drops below $\mathcal{O}(10^{-6})$ at 300~s, $\mathcal{O}(10^{-7})$ at 470~s, and $\mathcal{O}(10^{-8})$ at 525~s. In comparison to the dark sky, brightness is about 21.6 to $22.0~\mathrm{mag/arcsec^2}$ or $I/F \approx 6.2$ to $4.3 \times 10^{-9}$, which indicates that the curtain portion of the plume remains well above the dark-sky brightness floor.
\begin{figure*}[!t]
    \centering
    \includegraphics[width=0.9\textwidth]{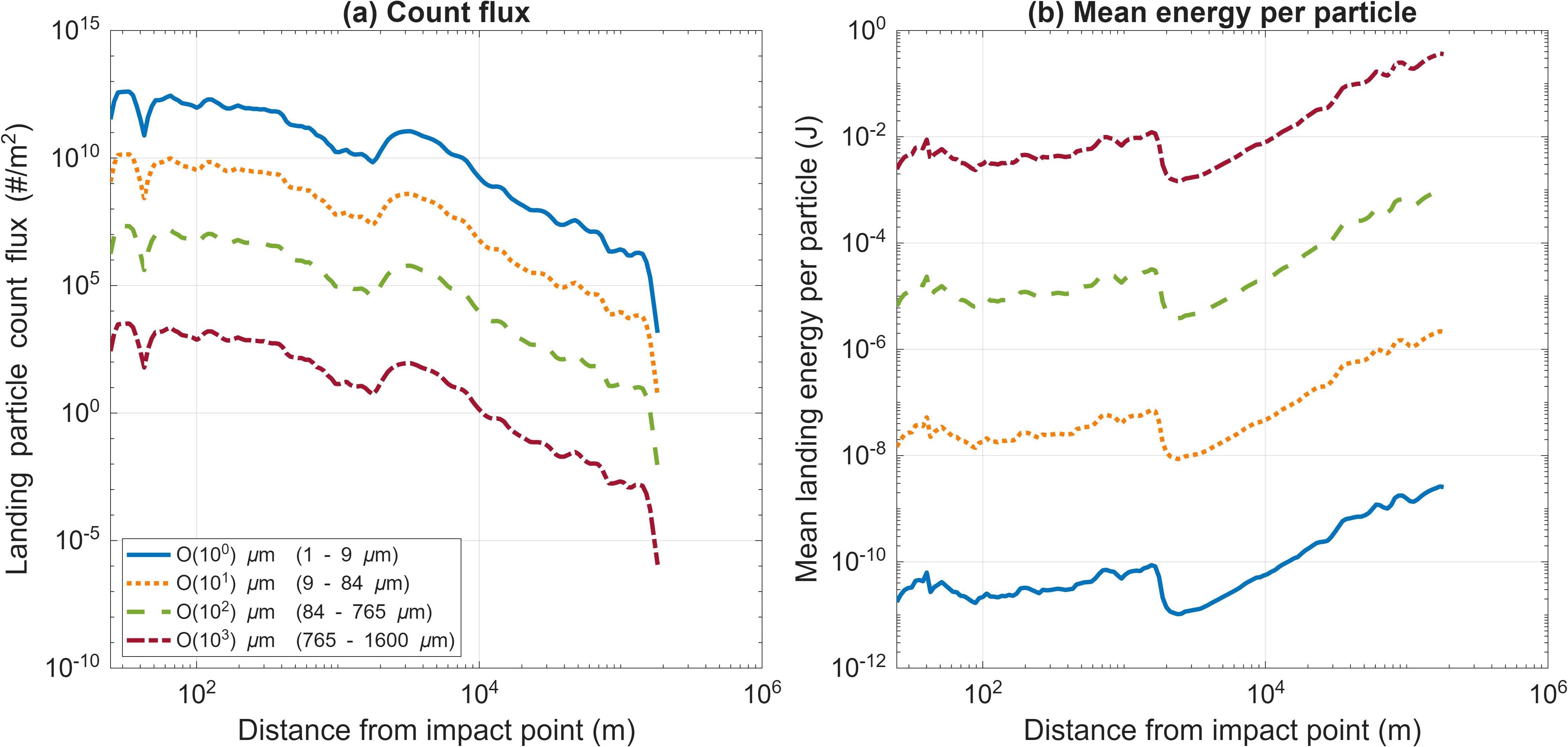}
    \caption{Landing particle (a) count flux and (b) mean kinetic energy per particle vs.\ distance from the impact point, separated into four grain-size bins (1 to 9~$\mu$m, 9 to 84~$\mu$m, 84 to 765~$\mu$m, 765 to 1600~$\mu$m).}
    \label{fig:landing_flux}
\end{figure*}
The fastest-moving ejecta from this impact orientation are expected to cover a ground-track radius of $\sim$183~km. We find that count flux decreases by $\sim$8 to 10 orders of magnitude from near the impact point to the edge of the ejecta field ($>10^{5}$~m), with the smallest grain sizes $\mathcal{O}(1)~\mu$m dominating the count flux by 3 to 4 orders of magnitude over the largest bin $\mathcal{O}(10^{3})~\mu$m at any given distance. Note that in our model, grains are created from Lagrangian tracer particles generated by iSALE, and therefore all grains created from a tracer move with the same velocity regardless of size. All four size bins therefore exhibit a similar non-monotonic structure consisting of an initial dip near the impact point, a rise to a local maximum around 1~km away, and then a decline toward the edge of the ejecta field. At the farthest resolved landing distances, $\mathcal{O}(10^{3})~\mu$m particles carry roughly 8 orders of magnitude more mean landing energy per particle than the smallest particle size bin $\mathcal{O}(1)~\mu$m at comparable distances. As expected, energy per particle increases with distance for all size bins. At the farthest landing distances resolved in our simulation, the $\mathcal{O}(10^{3})~\mu$m size particles have a landing count flux of roughly $10^{-5}~\mathrm{m^{-2}}$, which is equivalent to approximately one particle landing per $100{,}000~\mathrm{m^2}$ while carrying a mean kinetic energy of $\sim$0.36~J/particle. The full distance-resolved count flux and landing energy distributions are shown in Figure~\ref{fig:landing_flux}.

 We note that in our model, ejecta particle launch velocity is assigned per tracer particle independent of grain size; that is, a given tracer's velocity is not itself sorted by which grain-size bin its mass is subsequently split into. Instead, our size distribution is a mass-weighted split of each tracer's total ejecta, and not an independent ballistic population per grain size. This means that the largest grains reaching the farthest distances in our results are doing so simply because some tracers carrying that size fraction happen to have high launch velocity, not because of any physical sorting mechanism preferentially launching small particles fastest.
 \clearpage
\section{Conclusion}
The upcoming impact of the Falcon 9 upper stage near Einstein crater offers a rare opportunity to observe ejecta dynamics from a known artificial impactor under real lunar conditions. It is expected that under best conditions, the impact spot may be bright enough to see against the dark sky. However, an important caveat to this estimate is that our simulations model only an upright impact, a constraint of iSALE's 2D axisymmetric formulation. The ejecta dynamics for other impact orientations that Falcon 9 could have upon impact remain unknown. Thus, direct observation of the real event will offer a valuable opportunity to benchmark the predictions we have made with these simulations against actual measurements, and to assess how closely an upright-impact model captures the true impact geometry.

\section{Acknowledgements}
This work made use of the iSALE-2D shock physics code. We gratefully acknowledge the developers of iSALE-2D, including Gareth Collins, Kai Wünnemann, Dirk Elbeshausen, Tom Davison, Boris Ivanov and Jay Melosh. 

\section{Conflict of Interest Disclosure}
The authors declare there are no conflicts of interest for this manuscript.

\section{Contributions}
W.J. performed iSALE simulations, developed the Monte Carlo ballistics and radiometric post-processing, analyzed the results, and led the write-up of the manuscript. D.B.G. and P.L.V. supervised the research, contributed to the methodology, and provided edits to the manuscript. L.M.T. advised on the scattering and observability analysis. J.S. and A.M. contributed to the modeling methodology and provided edits to the manuscript. 

\section{Open Research}
The iSALE-2D shock physics code (Amsden et al., 1980; Collins et al., 2004; Wünnemann et al., 2006) is distributed to academic users upon registration at https://isale-code.github.io. Its license prohibits redistribution, including modified sources, so the tilted-cylinder object type implemented for this work cannot be shared directly. Rather, its geometric and material parameters are provided in the Supporting Information. The tracer particles data collected from the simulations, the Monte Carlo ballistic reconstruction, post-processing codes, and figure plotting codes are archived on Zenodo at https://doi.org/10.5281/zenodo.21521025 as a single zip file (2GB). Raw hydrocode output fields (jdata.dat files) are available from the corresponding author upon reasonable request.

\bibliographystyle{aa}
\bibliography{references}

%%%%%%%%%%%%%%%%%%%%%%%%%%%%%%%%%%%%%%%%%%%%%%%%%%%%%%%%%%%%%%%%
%% Appendices must be placed after   \end{thebibliography}
%% They will be placed automatically on a new page.
%%%%%%%%%%%%%%%%%%%%%%%%%%%%%%%%%%%%%%%%%%%%%%%%%%%%%%%%%%%%%%%%
% ============================================================
% SUPPORTING INFORMATION / APPENDIX
% Place this AFTER \end{thebibliography}
% ============================================================

\begin{appendix}
% The A&A template recommends one-column mode for appendices
% containing many full-width tables and figures.
\onecolumn
% ============================================================
\section{Impactor geometry setup}
\label{app:geometry}

\begin{table}[ht!]
\centering
\caption{Axisymmetric object definitions used to construct the Falcon~9
upper stage impactor geometry in iSALE-2D. The impactor was represented
using a combination of steel and void objects to reproduce the hollow
internal structure of the Falcon~9 upper stage. The Merlin engine nozzle
was modeled as a tilted cylinder (\texttt{CYL\_W\_TLT}) with a horizontal
radius of 54 cells, a vertical half-length of 2 cells, a tilt angle of
$113^{\circ}$, and a horizontal offset of 33 cells. Each object was assigned
independent horizontal and vertical geometric resolutions (CPPRH and CPPRV)
together with a vertical offset defining its position within the assembled
impactor geometry.}
\label{tab:geometry}

\begin{tabularx}{\textwidth}{@{}cllXccc@{}}
\hline\hline
Object ID & Material & Object type & Purpose / description
& CPPRH & CPPRV & Vertical offset \\
\hline
1  & Steel & CYLINDER
   & Interstage/adapter ring, upper section
   & 12 & 30 & 108 \\

2  & Steel & SPHEROID
   & Forward (LOX) tank dome
   & 73 & 40 & 168 \\

3  & Steel & CYLINDER
   & Main tank barrel wall
   & 73 & 176 & 208 \\

4  & Steel & SPHEROID
   & Common bulkhead (LOX/RP-1 tank divider)
   & 73 & 40 & 520 \\

5  & Void & CYLINDER
   & Main tank interior cavity
   & 69 & 176 & 208 \\

6  & Void & SPHEROID
   & Forward dome interior void
   & 69 & 36 & 524 \\

7  & Void & SPHEROID
   & Bulkhead interior void
   & 69 & 36 & 172 \\

8  & Steel & SPHEROID
   & Aft (RP-1) tank dome
   & 73 & 40 & 308 \\

9  & Void & SPHEROID
   & Aft dome interior void
   & 69 & 36 & 312 \\

10 & Void & CYLINDER
   & Engine bay cavity
   & 69 & 20 & 348 \\

11 & Steel & CYL\_W\_TLT
   & Merlin engine nozzle
   & 54 & 2 & 56 \\
\hline
\end{tabularx}
\end{table}

\FloatBarrier
% ============================================================
\section{Equation of state parameters}
\label{app:eos}

\begin{table}[ht!]
\centering
\caption{Tillotson equation of state parameters governing the thermodynamic
response of each material under shock compression and expansion. Lunar
regolith values are derived from basalt, following
\citet{benz1999catastrophic}, \citet{wojcicka2020seismic}, and
\citet{rajsic2021numerical}.}
\label{tab:eos}

\begin{tabular}{@{}llcc@{}}
\hline\hline
iSALE parameter & Description & Steel & Lunar regolith \\
\hline
TL\_RHO0
& Reference density (kg/m$^3$)
& $8.0 \times 10^{3}$
& $3.1 \times 10^{3}$ \\

TL\_CHEAT
& Specific heat capacity (J/kg/K)
& $4.20 \times 10^{2}$
& $1.0 \times 10^{3}$ \\

TL\_BULKA
& Bulk modulus (Pa)
& $1.17 \times 10^{11}$
& $1.93 \times 10^{10}$ \\

TL\_BULKB
& Tillotson $B$ constant (Pa)
& $5.5 \times 10^{10}$
& $2.93 \times 10^{10}$ \\

TL\_EZERO
& Tillotson $E_0$ constant (J/kg)
& $1.75 \times 10^{7}$
& $4.87 \times 10^{8}$ \\

TL\_THERMA
& Tillotson $a$ constant
& 0.55
& 0.50 \\

TL\_THERMB
& Tillotson $b$ constant
& 0.62
& 0.80 \\

TL\_ALPHA
& Tillotson $\alpha$ constant
& 5.0
& 5.0 \\

TL\_BETA
& Tillotson $\beta$ constant
& 5.0
& 5.0 \\

TL\_EIV
& SIE incipient vaporization (J/kg)
& $2.4 \times 10^{6}$
& $4.72 \times 10^{6}$ \\

TL\_ECV
& SIE complete vaporization (J/kg)
& $8.67 \times 10^{6}$
& $1.82 \times 10^{7}$ \\
\hline
\end{tabular}
\end{table}

\FloatBarrier
\newpage

% ============================================================
\section{Strength and damage model parameters}
\label{app:strength}

\begin{table}[ht!]
\centering
\caption{Strength and damage model parameters governing the mechanical
response of each material. The steel impactor is modeled using the
Johnson-Cook strength model, while the lunar regolith target is modeled
using the Drucker-Prager strength model. Parameters marked with a dash
(---) are not applicable to that material's constitutive model.}
\label{tab:strength}

\begin{tabular}{@{}llcc@{}}
\hline\hline
iSALE parameter & Description & Steel & Lunar regolith \\
\hline
STRMOD
& Strength model
& Johnson-Cook
& Drucker-Prager \\

DAMMOD
& Damage model
& None
& None \\

THSOFT
& Thermal softening
& Johnson-Cook
& Ohnaka \\

LDWEAK
& Low-density weakening
& None
& Polynomial \\

POIS
& Poisson's ratio
& 0.30
& 0.30 \\

YDAM0
& Cohesion (Pa)
& ---
& 10.0 \\

FRICDAM
& Internal friction
& ---
& 0.60 \\

YLIMDAM
& Limiting strength (Pa)
& ---
& $2.50 \times 10^{8}$ \\

YINT0
& Intact cohesion (Pa)
& ---
& $1.0 \times 10^{7}$ \\

FRICINT
& Intact friction
& ---
& 1.10 \\

YLIMINT
& Intact limiting strength (Pa)
& ---
& $2.50 \times 10^{9}$ \\

TMELT0
& Melt temperature (K)
& 1783.0
& 1327.0 \\

TFRAC
& Softening law
& ---
& 1.20 \\

ASIMON
& Simon coefficient $a$
& $4.402 \times 10^{10}$
& $6.0 \times 10^{9}$ \\

CSIMON
& Simon coefficient $c$
& 1.3978
& 3.0 \\

PMININ
& Minimum pressure (Pa)
& $-2.44 \times 10^{9}$
& --- \\

JC\_A
& Yield stress $A$ (Pa)
& $5.14 \times 10^{8}$
& --- \\

JC\_B
& Strain coefficient $B$ (Pa)
& $5.14 \times 10^{8}$
& --- \\

JC\_N
& Strain exponent $n$
& 0.508
& --- \\

JC\_C
& Strain-rate coefficient $c$
& 0.042
& --- \\

JC\_M
& Thermal softening exponent $m$
& 0.533
& --- \\

JC\_TREF
& Reference temperature (K)
& 293.0
& --- \\
\hline
\end{tabular}
\end{table}

\FloatBarrier

% ============================================================
\section{Material model summary}
\label{app:materials}

\begin{table}[ht!]
\centering
\caption{Constitutive model choices from the \#ISMAT input block. Steel
uses a Johnson-Cook strength model, while the five lunar regolith materials
(\texttt{lun\_rg1} to \texttt{lun\_rg5}) use a Drucker-Prager strength
model with W\"unnemann porosity compaction and Ohnaka thermal softening.
All materials use a Tillotson equation of state.}
\label{tab:materials}

\begin{tabular}{@{}lllllll@{}}
\hline\hline
Material & EOS & EOS type & Strength model
& Porosity model & Thermal softening & ALPHA0 \\
\hline
steel
& ststeel
& tillo
& JNCK (Johnson-Cook)
& WUNNEMA
& JNCK
& 27 \\

lun\_rg1
& lun\_reg
& tillo
& DRPR (Drucker-Prager)
& WUNNEMA
& OHNAKA
& 2.02 \\

lun\_rg2
& lun\_reg
& tillo
& DRPR (Drucker-Prager)
& WUNNEMA
& OHNAKA
& 1.76 \\

lun\_rg3
& lun\_reg
& tillo
& DRPR (Drucker-Prager)
& WUNNEMA
& OHNAKA
& 1.70 \\

lun\_rg4
& lun\_reg
& tillo
& DRPR (Drucker-Prager)
& WUNNEMA
& OHNAKA
& 1.67 \\

lun\_rg5
& lun\_reg
& tillo
& DRPR (Drucker-Prager)
& WUNNEMA
& OHNAKA
& 1.63 \\
\hline
\end{tabular}
\end{table}

\FloatBarrier
% ============================================================
\section{Regolith layer stratification}
\label{app:regolith}

\begin{figure}[ht!]
\centering
\includegraphics[width=0.82\textwidth]{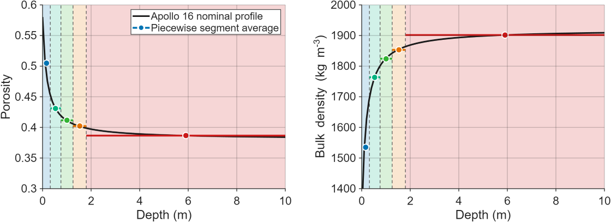}
\caption{Depth-dependent regolith porosity (left) and bulk density (right)
profiles based on the Apollo~16 nominal density law
$\rho(z)=1920\,(z+0.122)/(z+0.18)~\mathrm{kg\,m^{-3}}$ (black curve),
with a piecewise-constant approximation (colored step segments and markers)
used for the target-layer input. Vertical dashed lines mark segment
boundaries; shaded bands indicate the depth range of each segment, with
the outermost red segment extending to depths greater than 10~m.}
\label{fig:regolith_profile}
\end{figure}

\begin{table}[ht!]
\centering
\caption{Five regolith layers (\texttt{LAYNUM=5}) stacked by grid position,
each assigned a distinct initial porosity according to the
\citet{carrier1991physical} depth profile. Surface roughness is superimposed
as a sinusoid with an amplitude of 2 and wavelength of 55 grid cells.}
\label{tab:regolith}

\begin{tabular}{@{}cccc@{}}
\hline\hline
Layer \# & Material & Grid position (LAYPOS) & Porosity (ALPHA0) \\
\hline
1 & lun\_rg5 & 394 & 1.63 \\
2 & lun\_rg4 & 416 & 1.67 \\
3 & lun\_rg3 & 436 & 1.70 \\
4 & lun\_rg2 & 453 & 1.76 \\
5 & lun\_rg1 & 466 & 2.02 \\
\hline
\end{tabular}
\end{table}

\FloatBarrier
% ============================================================
\section{Apollo 16 particle size-frequency distribution}
\label{app:psd}

\begin{table}[ht!]
\centering
\caption{Apollo~16 particle size distribution
\citep{carrier1991physical} used to bin tracer mass and derive particle
count flux. The representative diameter of each bin is the arithmetic
mean of its lower and upper bounds.}
\label{tab:psd}

\begin{tabular}{@{}ccccc@{}}
\hline\hline
$d_{\mathrm{min}}$ ($\mu$m)
& $d_{\mathrm{max}}$ ($\mu$m)
& Mass fraction
& $d_{\mathrm{mean}}$ ($\mu$m)
& Relative mass \\
\hline
1   & 2    & 0.0402 & 1.5    & 0.0402 \\
2   & 4    & 0.0540 & 3      & 0.0540 \\
4   & 9    & 0.0723 & 6.5    & 0.0723 \\
9   & 19   & 0.1152 & 14     & 0.1152 \\
19  & 40   & 0.1719 & 29.5   & 0.1719 \\
40  & 84   & 0.1982 & 62     & 0.1982 \\
84  & 175  & 0.1517 & 129.5  & 0.1517 \\
175 & 366  & 0.1034 & 270.5  & 0.1034 \\
366 & 765  & 0.0743 & 565.5  & 0.0743 \\
765 & 1600 & 0.0188 & 1182.5 & 0.0188 \\
\hline
\end{tabular}
\end{table}

\FloatBarrier
% ============================================================
\section{Sun-target-observer geometry}
\label{app:geometry_sto}

\begin{table}[ht!]
\centering
\caption{Sun-target-observer geometry for the Moon at the predicted
impact epoch, obtained from JPL Horizons using the DE441 ephemeris.}
\label{tab:sto}

\begin{tabular}{@{}llc@{}}
\hline\hline
Quantity & Value & Units \\
\hline
Epoch (UT)
& 2026-Aug-05 06:34:00.0
& --- \\

Observer
& HST (spacecraft, body $-48$)
& --- \\

Range ($\Delta$)
& 0.0025
& AU \\

Range rate ($\dot{\Delta}$)
& 2.9
& km/s \\

S-O-T (solar elongation)
& 99.9 (/T, trailing)
& deg \\

S-T-O (phase angle)
& 80.0
& deg \\

Sky motion
& 208.5
& arcsec/min \\

Sky-motion position angle
& 290.2
& deg \\

RelVel-ANG
& 24.9
& deg \\
\hline
\end{tabular}
\end{table}

\FloatBarrier

\end{appendix}

\end{document}